\documentclass[twocolumn,epjc3]{svjour3}  

\journalname{Eur. Phys. J. A}

\usepackage{graphicx}
\usepackage{dcolumn}
\usepackage{bm}
\usepackage{multirow}
\usepackage{amssymb}
\usepackage{amsmath, bbm}
\usepackage{bbold}
\usepackage[figuresright]{rotating}
\usepackage{capt-of}
\usepackage{overpic}
\usepackage{subcaption}
\newcommand{\be}{\begin{eqnarray}}
\newcommand{\ee}{\end{eqnarray}}

\protected\def\lc{C}
\protected\def\ld{D}

\newcommand{\Ps}{P^{(\sigma)}}
\newcommand{\Pt}{P^{(\tau)}}

\newcommand\numberthis{\addtocounter{equation}{1}\tag{\theequation}}

\newlength{\feynwidth} 
\newlength{\feynwidthbig} 

\usepackage{xcolor}

\usepackage{hyperref}
\usepackage{graphicx}

\hypersetup{colorlinks=true,citecolor= blue}

\usepackage[english]{babel}
\usepackage[autolanguage]{numprint}

\usepackage{booktabs}
\usepackage{cite}

\begin{document}


\title{Benchmarking $\mathbf{\Lambda}$NN three-body forces and first predictions
for $A=3-5$ hypernuclei}
\author{Hoai Le\thanksref{addr1,e1}
\and Johann Haidenbauer\thanksref{addr1,e2}
\and Hiroyuki Kamada\thanksref{addr2,addr3,e3,e3s}
\and Michio  Kohno\thanksref{addr3,e4}
\and Ulf-G. Mei{\ss}ner\thanksref{addr4,addr1,addr6,e5}
\and Kazuya Miyagawa\thanksref{addr3,e6}
\and Andreas Nogga\thanksref{addr1,addr6,e7}
}
\thankstext{e1}{e-mail: h.le@fz-juelich.de}
\thankstext{e2}{e-mail: j.haidenbauer@fz-juelich.de}
\thankstext{e3}{e-mail: kamada@mns.kyutech.ac.jp}
\thankstext{e3s}{e-mail: kamada@rcnp.osaka-u.ac.jp}
\thankstext{e4}{e-mail: kohno@rcnp.osaka-u.ac.jp}
\thankstext{e5}{e-mail: meissner@hiskp.uni-bonn.de}
\thankstext{e6}{e-mail: miyagawa@rcnp.osaka-u.ac.jp}
\thankstext{e7}{e-mail: a.nogga@fz-juelich.de}

\institute{{Institute for Advanced Simulation (IAS-4), Forschungszentrum J\"ulich, 
D-52425 J\"ulich, Germany} \label{addr1}
\and
           {Department of Physics, Faculty of Engineering, Kyushu Institute of Technology, Kitakyushu 804-8550, Japan} \label{addr2}
             \and
           {Research Center for Nuclear Physics, Osaka University, Ibaraki 567-0047, Japan} \label{addr3}
           \and
           {Helmholtz-Institut~f\"{u}r~Strahlen-~und~Kernphysik~and~Bethe~Center~for~Theoretical~Physics,
~Universit\"{a}t~Bonn,~D-53115~Bonn,~Germany} \label{addr4}
           \and 
           {CASA, Forschungszentrum J\"ulich, D-52425 J\"ulich, Germany} \label{addr6}
}

\date{March 1, 2024}

\maketitle

\begin{abstract}
 Explicit expressions for the leading chiral hyperon-nucleon-nucleon three-body forces have been derived by Petschauer et al [Phys. Rev. C93.014001 (2016)]. An 
 important prerequisite for including these three-body forces in few- and many-body calculations is the accuracy and efficiency of their partial-wave decomposition.  A careful benchmark of the $\mathrm{\Lambda}$NN potential matrix elements, computed using two robust and efficient partial-wave decomposition methods, is presented. In addition, results of a first quantitative assessment 
for the contributions of $\mathrm{\Lambda}$NN forces to the separation energies in  $A=3-5$ hypernuclei are reported. 

\end{abstract}

\maketitle

\section{Introduction}
Few-nucleon systems have served as a crucial testing
ground for our understanding of  nucleon-nucleon (NN) and three-nucleon (3N) forces \cite{Nogga:2002kw,Epelbaum:2018ogq,Binder:2015mbz,LENPIC:2022cyu,Kalantar-Nayestanaki:2011rzs,Girlanda:2023znc,Lazauskas:2004uq,Deltuva:2017bia,Piarulli:2017dwd,Otsuka:2009cs}. 
In the course of this, due to the complexity of the computational treatment of few-body systems and the goal of achieving accurate predictions using realistic nuclear forces, it has become standard 
 to cross-compare results achieved with various methods and by independent research groups. Indeed, such benchmark studies have become an integral part of the advancement of 
microscopic few-nucleon calculations. For instance, in the past,
benchmark results have been produced for 
nucleon-deuteron (N-d) scattering~\cite{Cornelius:1990zz,Friar:1990zza},
for N-d breakup \cite{Friar:1995zz}, 
for the triton binding energy including $2\pi$ exchange
three-nucleon forces \cite{Nogga:2002qp}, 
for the four-nucleon (4N) bound state~\cite{Kamada:2001tv} and for 4N scattering \cite{Viviani:2011ax,Viviani:2016cww}.

Regarding strangeness nuclear physics realistic 
calculations of $\mathrm{\Lambda}$ hypernuclei
including the full complexity of the $\mathrm{\Lambda}$N-$\mathrm{\Sigma}$N
interaction were first presented in 
\cite{Miyagawa:1993rd,Miyagawa:1995sf} 
for the hypertriton and in 
\cite{Hiyama:2001zt,Nogga:2001ef,Nemura:2002fu}
for $^4_\mathrm{\Lambda}$H and $^4_\mathrm{\Lambda}$He. Both are
momentum-space calculations based on the Faddeev- and
Faddeev-Yakubovsky (FY) approaches, respectively. 
Very recently, the first calculations of the hypertriton
separation energy including chiral $\mathrm{\Lambda}$NN \\ three-body 
forces (3BFs) \cite{Petschauer:2016ho} have been published 
\cite{Kamada:2023txx,Kohno:2023xvh}.
Actual benchmark
studies for hypernuclei are however scarce. Over the years,
a diverse range of calculations employing various  methods 
\cite{Afnan:1989wb,Nemura:2002fu,Fujiwara:2007en,Garcilazo:2007ss,Wirth:2014apa,Wirth:2017bpw,Contessi:2018qnz,Frame:2020mvv,Le:2020zdu,Hildenbrand:2024ypw} have been carried out. However,
the elementary NN and hyperon-nucleon
(YN) potentials utilized as input in those calculations are very different, making a comprehensive 
comparison of the results not possible. 
On the other hand, an actual benchmark study
for few-body hypernuclei presented in Ref.~\cite{FerrariRuffino:2017otv}
relied on rather simple representations of the NN and
 YN interactions. 
Only lately, first elaborate benchmark results for 
$^4_\mathrm{\Lambda}$H \cite{Le:2020zdu} were reported, by comparing calculations based on the 
FY equations and the Jacobi no-core
shell model (Jacobi-NCSM),
for state-of-the-art NN and YN two-body interactions,
namely the so-called SMS NN potentials derived within
chiral effective field theory (EFT) \cite{Reinert:2017usi}
and YN interactions established likewise in chiral EFT 
\cite{Haidenbauer:2013oca,Haidenbauer:2019boi}.

With the present work we want to add a further benchmark
for $\mathrm{\Lambda}$ hypernuclei. 
Specifically, we provide a detailed comparison of the
calculations by Kamada, Kohno, and Miyagawa (KKM)
\cite{Kamada:2023txx,Kohno:2023xvh}
and the J\"ulich-Bonn Group (JBG) 
\cite{Le:2022ikc,Le:2023bfj} for the hypertriton 
including chiral 3BFs. The former calculation is performed
within the Faddeev approach while the latter utilizes 
the Jacobi-NCSM formalism. 
The motivation for our study originates from 
discrepancies in the contribution
of the $2\pi$ exchange $\mathrm{\Lambda}$NN force to the 
hypertriton separation energy observed between 
the KKM results \cite{Kamada:2023txx} and the preparatory 
calculations of JBG. In the course of clarifying them
\cite{KKM:2024},
it became clear that it would be rather useful to
provide an in-depth comparison of the results 
by the two groups, which does not only shed light
on the accuracy of the two methods but also allows for
an examination of the underlying technical and numerical 
aspects of such complex calculations. 
Clearly, such a detailed comparison is not
only indispensable for corroborating the outcome 
of the present three-body calculations, 
but it provides also a useful guideline for future 
calculations employing different few-body methods. 

The paper is organized as follow. In the following section, we briefly describe the two approaches for the partial-wave decomposition of the $\mathrm{\Lambda}$NN (and $\mathrm{\Sigma}$NN) potentials employed by KKM and JBG. A detailed comparison of the $\mathrm{\Lambda}$NN potential matrix elements in different partial-wave states are presented in Sect.~\ref{sec:compareLNN}.  In Sect.~\ref{sec:Eseparation_3_5} we discuss possible contributions of the chiral $\mathrm{\Lambda}$NN interaction to the separation energies in the $A=3-5$ hypernuclei  and we close with some concluding remarks. 

\section{Partial-wave decomposition of the chiral \texorpdfstring{$\mathrm{YNN}$}{YNN} forces}
\begin{figure*}
\centering
\hfill
\begin{subfigure}[t]{.3\textwidth}
\centering
\vspace{.3\baselineskip}
\begin{overpic}[scale=.6]{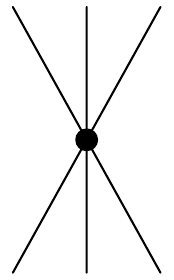}
\put(2,103){$l$}\put(27,103){$m$}\put(55,103){$n$}
\put(2,-10){$i$}\put(27,-10){$j$}\put(55,-10){$k$}
\put(0,-25){$A$}\put(25,-25){$B$}\put(53,-25){$C$}
\end{overpic}
\vspace{1.7\baselineskip}
\caption{}
\label{fig:cnt}

\end{subfigure}
\hfill
\begin{subfigure}[t]{.3\textwidth}
\centering
\vspace{.3\baselineskip}
\begin{overpic}[scale=.6]{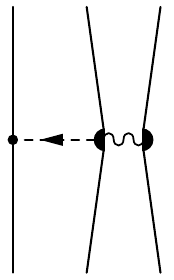}
\put(2,103){$l$}\put(27,103){$m$}\put(55,103){$n$}
\put(2,-10){$i$}\put(27,-10){$j$}\put(55,-10){$k$}
\put(0,-25){$A$}\put(25,-25){$B$}\put(53,-25){$C$}
\put(16,59){$\phi$}
\end{overpic}
\vspace{1.7\baselineskip}
\caption{ }
\label{fig:ome-gen1}
\end{subfigure}
\hfill
\begin{subfigure}[t]{.3\textwidth}
\centering
\vspace{.3\baselineskip}
\begin{overpic}[scale=.6]{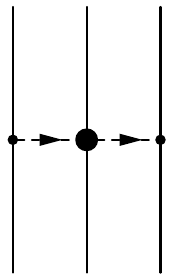}
\put(2,103){$l$}\put(27,103){$m$}\put(55,103){$n$}
\put(2,-10){$i$}\put(27,-10){$j$}\put(55,-10){$k$}
\put(0,-25){$A$}\put(25,-25){$B$}\put(53,-25){$C$}
\put(13,59){$\phi_1$} \put(39,59){$\phi_2$}
\end{overpic}
\vspace{1.7\baselineskip}
\caption{ }
\label{fig:tme-gen}
\end{subfigure}
\hfill\mbox{}
\caption{Generic $\mathrm{YNN} \rightarrow \mathrm{YNN}$ diagrams: (a) contact term, (b) one-meson exchange, (c) two-meson exchange.
The wiggly line symbolized the four-baryon contact vertex, to illustrate the baryon bilinears.}
\label{fig:diag_B1B2B3_B4B5B6}
\end{figure*}
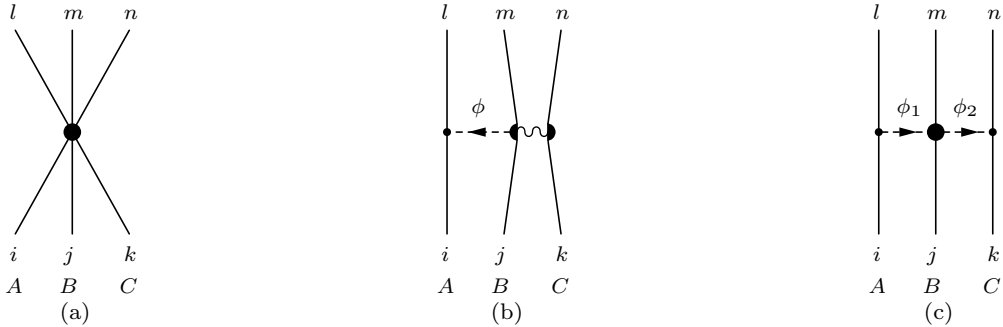

The generic contact, one- and two-meson exchange diagrams for the process $B_1 B_2 B_3  \rightarrow B_4B_5 B_6$,
appearing at next-to-next-to-leading order (N$^2$LO) in the
chiral expansion \cite{Petschauer:2016ho}, 
are shown in panels (a), (b) and (c) in 
Fig.~\ref{fig:diag_B1B2B3_B4B5B6}, respectively. The fully antisymmetrized contact $\mathrm{YNN}$ potential, obtained from the diagram (a) in Fig.~\ref{fig:diag_B1B2B3_B4B5B6} and all the permutations of the incoming $B_1 B_2 B_3$ and outgoing $B_4 B_5 B_6$ baryon states, is given by   \cite{Petschauer:2016ho,dissertation_stf}
\begin{align}\label{eq:ctterm}
\begin{split}
V_\mathrm{ct}  & = -\big [ N^1_{\substack{456\\123}}  + N^2_{\substack{456 \\123}}\vec{\sigma}_A . \vec{\sigma}_B +   N^3_{\substack{456 \\123}}\vec{\sigma}_A . \vec{\sigma}_C  +  N^4_{\substack{456 \\123}}\vec{\sigma}_B . \vec{\sigma}_C  \\[3pt]
 &\qquad  +  N^5_{\substack{456 \\123}}  i \vec{\sigma}_A . (\vec{\sigma}_B \times \vec{\sigma}_C)  \big] ,
\end{split}
\end{align}
where $N^{i}_{\substack{456 \\123}}$ are appropriately antisymmetrized
combinations of the 18 LECs defined in Eq.~(18) of Ref.~\cite{Petschauer:2016ho}. The one-meson exchange 
 potential corresponding to the master diagram (b) in 
 Fig.~\ref{fig:diag_B1B2B3_B4B5B6} reads,
\begin{align}\label{eq:1meterm}
\begin{split}
V_\mathrm{1me}  & =  \frac{1}{2f^2_{0}} \frac{\vec{\sigma}_A .\vec{q}_{li}}{{\vec{q}_{li}}^{\: 2} + m^2_{\phi}}\big [ N_1  \vec{\sigma}_C . \vec{q}_{li} + N_2 i (\vec{\sigma}_B \times \vec{\sigma}_C) \vec{q}_{li}  \big] ,
\end{split}
\end{align}
with $\vec{q}_{li} =\vec{p}_l - \vec{p}_i$ the transferred momentum. Explicit expressions for the constants $N_1, N_2$ are given by Eq.~(30) in Ref.~\cite{Petschauer:2016ho}. Based on the general expression in Eq.~(\ref{eq:1meterm}), the  antisymmetrized one-meson exchange $\mathrm{B_1 B_2 B_3} \rightarrow \mathrm{B_4 B_5 B_6 }$ potential  can  be obtained by summing up for each exchange  meson $\phi$ the 36 permutations of the initial and final baryons. Finally, the two-meson exchange diagram (c) yields 
\begin{align}\label{eq:2meterm}
\begin{split}
V_\mathrm{2me} = & -\frac{1}{4f_0^4} \frac{\vec{\sigma}_A. \vec{q}_{li} \, \vec{\sigma}_C . \vec{q}_{nk}}{ (\vec{q}^{\: 2}_{li} + m^2_{\phi_1}) (\vec{q}^{\: 2}_{nk} +m^2_{\phi_2})}\\[3pt]
& \qquad \times [ N^{\prime}_1  + N^{\prime}_2 \vec{q}_{li} . \vec{q}_{nk}  + N^{\prime}_3 i (\vec{q}_{li} \times \vec{q}_{nk}) .\vec{\sigma}_B] .
\end{split}
\end{align}
The constants $N^{\prime}_{1,2,3}$ are defined  in Eq.~(34) in Ref.~\cite{Petschauer:2016ho}. Similarly, summing up Eq.~(\ref{eq:2meterm}) for all the 18 permutations\footnote{The contribution from those permutations that yield identical results to the diagram in (c) is already included in Eq.~(\ref{eq:2meterm}), which explains for the factor of 18 instead of 36.} of the initial and final  baryon states and all  possible exchanged mesons,  one  obtains the general
  antisymmetrized two-meson exchange YNN potential. Note that, 
  in the calculations by JBG, all the coefficients $N^{i}_{\substack{456 \\123}}$, $N_{i}$ and $N^{\prime}_{i}$ in Eqs.~(\ref{eq:ctterm}-\ref{eq:2meterm}) have been evaluated as functions of the involving LECs using {\it Mathematica}. 
  
For the case of   $\mathrm{\Lambda NN} \rightarrow \mathrm{\Lambda NN}$ 3BFs that involve only $\pi$-meson exchanges, 
 the expressions for the $V^{\mathrm{\Lambda NN}}$ potentials 
 in Eqs.~(\ref{eq:ctterm}-\ref{eq:2meterm}) can be simplified significantly \cite{Petschauer:2016ho},
 \begin{align*}\label{eq:LNNct}
V^\mathrm{\Lambda NN}_\mathrm{ct} ={}
& \phantom{{}+{}} \lc'_1\ (\mathbbm1 - \vec\sigma_2\cdot\vec\sigma_3 ) ( 3 + \vec\tau_2\cdot\vec\tau_3 ) \\
& + \lc'_2\ \vec\sigma_1\cdot(\vec\sigma_2+\vec\sigma_3)\,(\mathbbm1 - \vec\tau_2\cdot\vec\tau_3) \\
& + \lc'_3\ (3 + \vec\sigma_2\cdot\vec\sigma_3 ) ( \mathbbm1 - \vec\tau_2\cdot\vec\tau_3 ) \,, \numberthis
\end{align*}

\begin{align*} \label{eq:LNNope}
V^\mathrm{\Lambda NN}_\mathrm{1\pi} =&{} -\frac{g_A}{2f_0^2} \,\bigg( \\
&\frac{\vec\sigma_2\cdot\vec q_{52}}{\vec q_{52}^{\,2}+m_\pi^2} \vec\tau_2\cdot\vec\tau_3
\Big[  (\ld'_1\vec\sigma_1+\ld'_2\vec\sigma_3)\cdot\vec q_{52} \Big] \\
&+\frac{\vec\sigma_3\cdot\vec q_{63}}{\vec q_{63}^{\,2}+m_\pi^2} \vec\tau_2\cdot\vec\tau_3
\Big[  (\ld'_1\vec\sigma_1+\ld'_2\vec\sigma_2)\cdot\vec q_{63} \Big] \\
&+\Ps_{23}\Pt_{23} \Ps_{13}\frac{\vec\sigma_2\cdot\vec q_{62}}{\vec q_{62}^{\,2}+m_\pi^2} \vec\tau_2\cdot\vec\tau_3 \\
&\qquad\times\Big[  -\frac{\ld'_1+\ld'_2}2 (\vec\sigma_1+\vec\sigma_3)\cdot\vec q_{62} \\
&\qquad\qquad+ \frac{\ld'_1-\ld'_2}2\,\mathrm i\,(\vec\sigma_3\times\vec\sigma_1)\cdot\vec q_{62}  \Big] \\
&+\Ps_{23}\Pt_{23} \Ps_{12}\frac{\vec\sigma_3\cdot\vec q_{53}}{\vec q_{53}^{\,2}+m_\pi^2} \vec\tau_2\cdot\vec\tau_3 \\
&\qquad\times\Big[  -\frac{\ld'_1+\ld'_2}2 (\vec\sigma_1+\vec\sigma_2)\cdot\vec q_{53} \\
&\qquad\qquad- \frac{\ld'_1-\ld'_2}2\,\mathrm i\,(\vec\sigma_1\times\vec\sigma_2)\cdot\vec q_{53}  \Big]
\bigg)\,, \numberthis
\end{align*}
and,
\begin{align*} \label{eq:LNNtpe}
&V^\mathrm{\Lambda NN}_\mathrm{2\pi} ={}
\frac{g_A^2}{3f_0^4}
\frac{\vec\sigma_3\cdot\vec q_{63}\ \vec\sigma_2\cdot\vec q_{52}}{(\vec q_{63}^{\,2}+m_{\pi}^2)(\vec q_{52}^{\,2}+m_{\pi}^2)} \vec\tau_2\cdot\vec\tau_3 \\
&\qquad \ \times \Big( -(3 b_0 + b_D) m_\pi^2      +      (2 b_2 + 3 b_4)      \,\vec q_{63}\cdot\vec q_{52}\Big) \\
&\quad- \Ps_{23}\Pt	_{23} \frac{g_A^2}{3f_0^4}
\frac{\vec\sigma_3\cdot\vec q_{53}\ \vec\sigma_2\cdot\vec q_{62}}{(\vec q_{53}^{\,2}+m_{\pi}^2)(\vec q_{62}^{\,2}+m_{\pi}^2)} \vec\tau_2\cdot\vec\tau_3 \\
&\qquad \ \times \Big( -(3 b_0 + b_D) m_\pi^2      +      (2 b_2 + 3 b_4)      \,\vec q_{53}\cdot\vec q_{62}\Big) \,. \numberthis
\end{align*}
Here, the $C_i', D_i', b_i$ are low-energy constants (LECs), the latter can in principle be fixed from the octet baryon masses 
and three-flavor meson-baryon scattering~\cite{Mai:2009ce}.
Note that, when the potentials in Eqs.~(\ref{eq:ctterm}-\ref{eq:LNNtpe}) are applied to basis wave functions $\mathrm{YNN}$ ($\mathrm{\Lambda NN}$) for which the two-nucleon states are antisymmetric, a scaling factor of $\frac{1}{2}$ \cite{Kohno:2022bsq,Kamada:2023txx,Kohno:2023xvh} is required.

In order to include the above $\mathrm{YNN}$ ($\mathrm{\Lambda NN}$) interactions in few- and many-body hypernuclear calculations,  efficient and accurate methods for the partial-wave decomposition of these potentials are of  importance.  Therefore, in this study, we want to benchmark the chiral potential matrix elements $V^{\mathrm{\Lambda NN}}$ evaluated using two different partial-wave decomposition methods. In the first approach, referred to as lPWD, the locality of the chiral $\mathrm{\Lambda NN}$ potentials in Eqs.~(\ref{eq:LNNct}-\ref{eq:LNNtpe}) is explicitly exploited so that the eight-fold integration  over the angles  of the incoming and outgoing momenta can be reduced to a two-fold integration, which in turn can significantly speed up the generation of the 3BF matrix elements.
This method has initially been applied to  the local chiral 3NFs up to $\mathrm{N^3LO}$ 
 by Hebeler et al \cite{Hebeler:2015wxa},
and recently extended by KKM \cite{Kohno:2022bsq} to compute  the partial-wave
decomposition matrix elements of the chiral $\mathrm{\Lambda NN}$ 3BF
at $\mathrm{N^2LO}$ based on Eqs.~(\ref{eq:LNNct}-\ref{eq:LNNtpe}). In the method
of KKM, the $\mathrm{\Lambda NN}$ interactions are rewritten in the tensor product
form by separating the spin and angular-momentum parts and a convenient expression
in a form similar to the Wigner-Eckart theorem is derived for the matrix element of
the angle-dependent term. For more details, one can refer to \cite{Kohno:2022bsq}.
In the second approach, utilized by JBG and referred to 
as aPWD, 
the technique introduced by Skibinski et al. in Ref.~\cite{Skibinski:2011db} is employed
to automatically perform the  partial-wave decomposition of both $\mathrm{\Lambda NN}$ and $\mathrm{\Sigma NN}$ potentials using the general expressions in Eqs.~(\ref{eq:ctterm}-\ref{eq:2meterm}).

In general, the three-body $\mathrm{YNN}$ partial-wave states $|p_{12} q_3\alpha_{\mathrm{YNN}}\rangle$ with the total angular momentum $J$ and  total isospin $T$   in  $jj$-coupling can be constructed  as follows
\begin{align}\label{eq:jjcoupling}
\begin{split}
&|p_{12} q_{3}\alpha_{\mathrm{YNN}} \rangle  \\
& =| p_{12} q_3 (l_{12} s_{12})j_{12} (l_3 \frac{1}{2})I_3  (j_{12} I_3) J M_J, (t_{12} t_{\mathrm{Y}}) T M_T \rangle,
\end{split}
\end{align}
where $p_{12}$ and $q_3$ are the relative Jacobi momenta between two nucleons and between the center-of-mass of two nucleons and the hyperon, respectively, and $\alpha_{\mathrm{YNN}}$ denotes a set of discrete quantum numbers characterizing the state. In the first step of the aPWD, the 3BF $\mathrm{YNN}$ matrix elements are calculated 
in the partial-wave state in $LS$-coupling,
$|p_{12} q_3 \beta_{\mathrm{YNN}}\rangle$,
\begin{align}\label{eq:lscoupling}
\begin{split}
&|p_{12} q_{3}\beta_{\mathrm{YNN}} \rangle  \\
& =| p_{12} q_3 (l_{12} l_3) L (s_{12} \frac{1}{2})S  (L,S) J M_{J}, (t_{12} t_{\mathrm{Y}}) T M_{T} \rangle.
\end{split}
\end{align}
The $LS$-coupling representation $|p_{12} q_{3}\beta_{\mathrm{YNN}} \rangle$ is related to the basis $ |p_{12} q_{3}\alpha_{\mathrm{YNN}}\rangle$ in Eq.~(\ref{eq:jjcoupling}) simply via a $9j$ symbol and  Clebsch-Gordan coefficients \cite{Glockle:1983sp}. In the basis $|p_{12} q_{3}\beta_{\mathrm{YNN}} \rangle$, the 3BF $\mathrm{YNN}$ matrix elements can be expressed as
\begin{align}\label{eq:YNNmatrix}
\begin{split}
& \langle p_{12}^{\prime} q_{3}^{\prime} \beta^{\prime}_{\mathrm{YNN}} | V_{\mathrm{YNN}} |
p_{12} q_{3} \beta_{\mathrm{YNN}} \rangle=\\
&  \int d\hat{p}_{12}^{\prime} \int d\hat{q}_3^{\prime}  \int d\hat{p}_{12} \int d\hat{q}_3\\
&  \sum_{m_{L^{\prime}}}  C(L^{\prime} S^{\prime} J^{\prime}; m_{L^{\prime}}, M_{J^{\prime}} -m_{L^{\prime}}, M_{J^{\prime}}) \mathcal{Y}^{* L^{\prime}, m_{L^{\prime}}}_{l^{\prime}_{12} l^{\prime}_{3}}(\hat{p}_{12}^{\prime} \hat{q}_3^{\prime})\\
&  \sum_{m_{L}}  C(L S J; m_{L}, M_{J} -m_{L}, M_{J}) \mathcal{Y}^{L, m_{L}}_{l_{12} l_{3}}(\hat{p}_{12} \hat{q}_3)\\
& \big \langle p^{\prime}_{12} q^{\prime}_{3}\,(s^{\prime}_{12} \frac{1}{2})S^{\prime} M_{J} - m_{L^{\prime}} (t^{\prime}_{12}, t_{Y^{\prime}}) T^{\prime} M_{T} | V^{\mathrm{YNN}} |  \\
& \quad  p_{12} q_{3} \,(s_{12} \frac{1}{2})S M_{J} - m_{L} (t_{12}, t_{Y}) T M_{T} \rangle,
\end{split}
\end{align}
where
\begin{align}
\begin{split}
\mathcal{Y}_{l_{12}l_3}^{L m_{L}}(\hat{p}_{12} \hat{q}_3) =& \!\!\!\sum_{m_{l_{12}}=-l_{12}}^{l_{12}}\!\!\!\! C(l_{12}, l_3, L; m_{l_{12}}, m_L -m_{l_{12}},m_L )\\
& \quad \times Y_{l_{12},m_{l_{12}}}(\hat{p}_{12})  Y_{l_3, m_L - m_{l_{12}} }(\hat{q}_{3}).
\end{split}
\end{align}
The matrix elements in the spin- and isospin-spaces in Eq.~(\ref{eq:YNNmatrix}), $ \big \langle p^{\prime}_{12} q^{\prime}_{3}\,(s^{\prime}_{12} \frac{1}{2})S^{\prime} M_{S^{\prime}}  (t^{\prime}_{12}, t_{Y^{\prime}}) T^{\prime} M_{T} | V^{\mathrm{YNN}} | $ \newline $   p_{12} q_{3} \,(s_{12} \frac{1}{2})S M_{S} (t_{12}, t_{Y}) T M_{T} \rangle $, depend on the mo\-men\-ta, spin- and isospin-quantum numbers of the incoming and outgoing states.  They can  be computed in analytic form as a function of the momenta $p_{12}, q_{3}$ and $p^{\prime}_{12}, q^{\prime}_3$
for all combinations  of spin and isospin- 
quantum numbers utilizing a software for symbolic calculations such as {\it Maple} (in our case) or {\it Mathematica} \cite{Skibinski:2011db}. This symbolic software also allows an automatic generation of a FORTRAN 
code for these matrix elements, so that the multifold  integration over the  angular part in Eq.~(\ref{eq:YNNmatrix}) can efficiently be  calculated numerically  using a FORTRAN program. Furthermore, given that the 3BFs $V^{\mathrm{YNN}}$ is rotationally invariant, the matrix elements in Eq.~(\ref{eq:YNNmatrix})  vanish unless $J= J^{\prime}$ and $M_{J} = M_{J^{\prime}}$, and in addition, they do not depend on the magnetic quantum number $M_{J}$, hence
\begin{align}\label{eq:YNNmatrix2}
\begin{split}
& \langle p_{12}^{\prime} q_{3}^{\prime} \beta^{\prime}_{\mathrm{YNN}} | V_{\mathrm{YNN}} |
p_{12} q_{3} \beta_{\mathrm{YNN}} \rangle=\\
& \int d\hat{p}_{12}^{\prime} \int d\hat{q}_3^{\prime}  \int d\hat{p}_{12} \int d\hat{q}_3\, \frac{1}{2J+1} \sum_{m_{J}=-J}^{J}\\
&  \sum_{m_{L^{\prime}}}  C(L^{\prime} S^{\prime} J; m_{L^{\prime}}, M_{J} -m_{L^{\prime}}, M_{J}) \mathcal{Y}^{* L^{\prime}, m_{L^{\prime}}}_{l^{\prime}_{12} l^{\prime}_{3}}(\hat{p}_{12}^{\prime} \hat{q}_3^{\prime})\\
&  \sum_{m_{L}}  C(L S J; m_{L}, M_{J} -m_{L}, M_{J}) \mathcal{Y}^{ L, m_{L}}_{l_{12} l_{3}}(\hat{p}_{12} \hat{q}_3)\\
& \big \langle p^{\prime}_{12} q^{\prime}_{3}\,(s^{\prime}_{12} \frac{1}{2})S^{\prime} M_{J} - m_{L^{\prime}} (t^{\prime}_{12}, t_{Y^{\prime}}) T^{\prime} M_{T} | V^{\mathrm{YNN}} |  \\
& \quad  p_{12} q_{3} \,(s_{12} \frac{1}{2})S M_{J} - m_{L} (t_{12}, t_{Y}) T M_{T} \rangle.
\end{split}
\end{align}
 Since the integrand in Eq.~(\ref{eq:YNNmatrix2}) is a scalar, one can therefore freely   chose the directions of the momenta, say $p^{\prime}_{12}$ and   $q^{\prime}_{q3}$  such that $\vec{p}_{12} = (0,0,p_{12})$ and $\phi_{q_3} =0$. As a consequence, the eight-fold integration in Eq.~(\ref{eq:YNNmatrix2}) can be  effectively reduced to a five-fold integration \cite{Skibinski:2011db},
 $$
 \int d\hat{p}_{12}^{\prime}  \int d\hat{q}_{3}^{\prime} \int d\hat{p}_{12} \int d\hat{q}_3 \rightarrow  \int d \theta_{q_3}\int d\hat{p}_{12} \int d\hat{q}_3~,
 $$
 which, in turn, can lead to a significant speed-up of the 
 generation of the 3BF matrix-elements. Once the 3BF matrix elements in the $LS$-representation are known, the recoupling to  the $jj$-basis, $ \langle p_{12}^{\prime} q_{3}^{\prime} \alpha^{\prime}_{\mathrm{YNN}} | V_{\mathrm{YNN}} |\\
p_{12} q_{3} \alpha_{\mathrm{YNN}} \rangle $, can easily be done \cite{Glockle:1983sp}. In addition, since we assume that the 3BF YNN is charge independent, it is therefore sufficient to compute the matrix elements in Eq.~(\ref{eq:YNNmatrix2}) for a specific value of $m_T$, say $m_{T} =0$.
\section{Benchmarking \texorpdfstring{$\mathrm{\Lambda}$NN}{LambdaNN} matrix elements}\label{sec:compareLNN}
We are now in a position to benchmark the 3BF matrix elements comput\-ed using the two different partial-wave decomposition approaches described in the previous section. Since the lPWD method has only been  implemented for the   $\mathrm{\Lambda NN}$ potential, we will focus on  comparing  the $\mathrm{\Lambda NN}$ potential matrix elements and turn off the $\mathrm{\Sigma}$ components in the aPWD approach also for the binding energy calculations discussed later. In Table~\ref{tab:YNNstates}, we list the quantum numbers of the $\alpha_{\mathrm{\Lambda NN}}$ states with positive parity and  the total angular momentum  and  isospin of  $(J^{\pi}, T) = (1/2^+,0)$ and $(3/2^+,0)$ which have been selected for benchmarking. The $2\pi$-exchange $\mathrm{\Lambda NN}$ matrix elements, $\langle p^{\prime}_{12} q^{\prime}_{3} \alpha^{\prime}_{\mathrm{\Lambda NN}} | V_{2\pi} | p_{12} q_3 \alpha_{\mathrm{\Lambda NN}} \rangle$, comput\-ed at fixed Jacobi momenta \\ $p^{\prime}_{12} = p_{12} =q_{3} = 0.205507$~fm\textsuperscript{-1} and $q^{\prime}_{3} = 0.306967$~fm\textsuperscript{-1} are presented in Table~\ref{tab:compare_VYNNmatrix2pe}. The sub-leading meson-baryon LECs \cite{Petschauer:2016ho},
appearing in $V^\mathrm{\Lambda NN}_\mathrm{2\pi}$,
 have been set to $3b_0 + b_D  = 0$   and $2b_2 + 3b_4 = -3.0\times  10^{-3} $~MeV\textsuperscript{-1}. One can clearly observe almost perfect agreement between the aPWD and lPWD $2\pi$-exchange $\mathrm{\Lambda NN}$ matrix elements.
 
 \begin{table*}[tbp]
\renewcommand{\arraystretch}{1.6}
\begin{center}
  \setlength{\tabcolsep}{0.2cm}
\begin{tabular}{ r  r   r  r r r  r r }
\hline
$(\mathrm{J}^{\pi},T)$&   $\alpha_{\mathrm{\Lambda NN}}$ & $l_{\mathrm{12}}$ & $s_{\mathrm{12}}$ & $J_{\mathrm{12}}$ & $t_{\mathrm{12}}$ & $l_{\mathrm{\Lambda}}$  & $2I_{\mathrm{\Lambda}}$  \\
\hline
\multirow{3}{*}{\shortstack[l]{$(1/2^+,0)$ }}   & 1 & 0 &  1  & 1 & 0 & 0 & 1 \\
&    2 & 2 &  1  & 1 & 0 & 0 & 1\\
&    3 & 1 &  0  & 1 & 0 & 1 & 1\\
\hline
\hline
\multirow{3}{*}{\shortstack[l]{($3/2^{+},0)$ }}   & 1 & 0 &  1  & 1 & 0 & 0 & 1 \\
&    2 & 2 &  1  & 1 & 0 & 0 & 1\\
&    3 & 1 &  0  & 1 & 0 &1 & 1\\
 \hline
  \end{tabular}
\end{center}
\caption{Quantum numbers of the first three  $\mathrm{\Lambda}$NN partial-wave states for the two selected
         $J^\pi$ and $T$.}
\label{tab:YNNstates}
\renewcommand{\arraystretch}{1.4}
\end{table*}

\begin{table*}[tbp]
\renewcommand{\arraystretch}{1.6}
\begin{center}
  \setlength{\tabcolsep}{0.2cm}
\begin{tabular}{ r r  r   r   r   r r  r }
\hline
&   & \multicolumn{3}{c}{\shortstack[l]{J=1/$2^{+}$,  T=0 }}   &   \multicolumn{3}{c}{\shortstack[l]{J=3/$2^{+}$, T=0 }}   \\
 $\alpha^{\prime}_\mathrm{\Lambda NN}$ & $\alpha_\mathrm{\Lambda NN}$ &   aPWD    & lPWD   & diff [\%] &     aPWD  & lPWD  & diff [\%]  \\
 \hline
 \hline
  1   &   1  &       0.211808E-03   &  0.211795E-03  & 0.01    &   0.211818E-03  &  0.211795E-03 & 0.01\\
  2   &   1    &    0.488366E-03  &  0.488674E-03  &  0.06   &  0.488367E-03 & 0.488674E-03   &0.06\\
  3   &   1    &     0.200297E-03  & 0.200317E-03  &  0.01   &  -0.100145E-03 & -0.100158E-03  & 0.01\\
  1   &   2    &     0.488614E-03    &  0.488674E-03  &   0.01  &  0.488511E-03 & 0.488674E-03   & 0.03\\
  2   &   2    &    -0.781242E-04   & -0.781013E-04  &   0.03  & -0.781352E-04 & -0.781013E-04   &0.04\\
  3   &   2    &     0.504514E-04  & 0.504487E-04  &      0.01 &  -0.252244E-04 & -0.252244E-04  & \\
  1   &   3    &    0.112725E-03  & 0.112723E-03  &   0.002  & -0.563600E-04 & -0.563617E-04  & 0.03\\
  2   &   3    &     0.341903E-04  &  0.341810E-04 &   0.03  &   -0.170948E-04  &  -0.170905E-04 & 0.02\\
  3   &   3    &    0.779062E-04  & 0.779012E-04  &  0.01   &   0.779025E-04 & 0.779012E-04   & 0.02\\
\hline                             
  \end{tabular}
\end{center}
\caption{ 2$\pi$-exchange $\mathrm{\Lambda NN}$ matrix elements   $\langle p^{\prime}_{12} q^{\prime}_{3} \alpha^{\prime}_\mathrm{\Lambda NN} | V_{\mathrm{2\pi}} | p_{12} q_3 \alpha_\mathrm{\Lambda NN} \rangle$ in fm\textsuperscript{5},  computed with the automatic 
 partial-wave decomposition (aPWD) and the approach that exploits the locality of the chiral YNN interaction (lPWD). The incoming and outgoing momenta are fixed to $p^{\prime}_{12} = p_{12} = q_3 = 0.205507 $ fm\textsuperscript{-1} and $q^{\prime}_3 =0.306967$ fm\textsuperscript{-1}.
 The sub-leading meson-baryon LECs \cite{Petschauer:2016ho}
 are set to $3b_0 + b_D  = 0$   and $2b_2 + 3b_4 = -3.0\times  10^{-3} $ MeV\textsuperscript{-1} .  }
\label{tab:compare_VYNNmatrix2pe}
\end{table*}
 \begin{table*}[tbp]
\renewcommand{\arraystretch}{1.6}
\begin{center}
  \setlength{\tabcolsep}{0.2cm}
\begin{tabular}{ r r  r   r   r   r r  r }
\hline
 \multicolumn{2}{c}{J=1/$2^{+}$,  T=0 }   &   \multicolumn{3}{c}{ $V_{1\pi}$}   &  \multicolumn{3}{c}{ $V_{ct} $  }  \\
 $\alpha^{\prime}_\mathrm{\Lambda NN}$ & $\alpha_\mathrm{\Lambda NN}$ &    aPWD    & lPWD   & diff [\%] &     aPWD  & lPWD  & diff [\%]  \\
 \hline
 \hline
  1   &   1  &       0.166474E-02    &  0.167123E-02  &  0.4  &  0.379766E-02  &  0.380185E-02 & 0.1\\
  2   &   1    &     0.156132E-02  & 0.156852E-02    & 0.4   &   0.0  &  0.0 & \\
  3   &   1    &     -0.271857E-12  & 0.0  &     &  0.0 & 0.0 &    \\
  1   &   2    &     0.156197E-02    &0.156852E-02&   0.4  &  0.0 & 0.0 & \\
  2   &   2    &    0.479602E-02   & 0.481549E-02  &   0.4  &   -0.249311E-08 & -0.906432E-10   & \\
  3   &   2    &     0.476842E-13   &  0.0 &       &  -0.181852E-13 &0.0  &  \\
  1   &   3    &     0.317498E-13   & 0.0  &     & 0.0  & 0.0 & \\
  2   &   3    &     0.857880E-13  & 0.0  &     &   0.0  &0.0  &  \\
  3   &   3    &     0.503546E-04  & 0.505587E-04   &  0.4   &   -0.0 & 0.603799E-19 & \\ 
\hline                           
  \end{tabular}
\end{center}
\caption{ Contact and  1$\pi$-exchange $\mathrm{\Lambda NN}$ matrix elements  in fm\textsuperscript{5}, computed with the automatic 
 partial-wave decomposition (aPWD) and the approach that exploits the locality of the chiral YNN interaction (lPWD). The incoming and outgoing momenta are fixed to $p^{\prime}_{12} = p_{12} = q_3 = 0.205507 $ fm\textsuperscript{-1} and $q^{\prime}_3 =0.306967$ fm\textsuperscript{-1}. LECs are set to  $D^{\prime}_1 =0$, $D^{\prime}_2 = \frac{2 C }{9 f_0^2 \Delta}  = 
 0.6268$~fm$^3$
 with $C=3/4\, g_A$ and $f_0 = 93$  MeV, $\Delta =300$ MeV, and $C^{\prime}_2=0$,
 $C^{\prime}_1 = C^{\prime}_3 = \frac{1}{72 f_0^4 \Delta} 
 = 0.1852$~fm$^5$.
 }
\label{tab:compare_VYNNmatrix1pect}
\end{table*}
\begin{figure}[tbp] 
      \begin{center}
        { \includegraphics[width=0.5\textwidth]{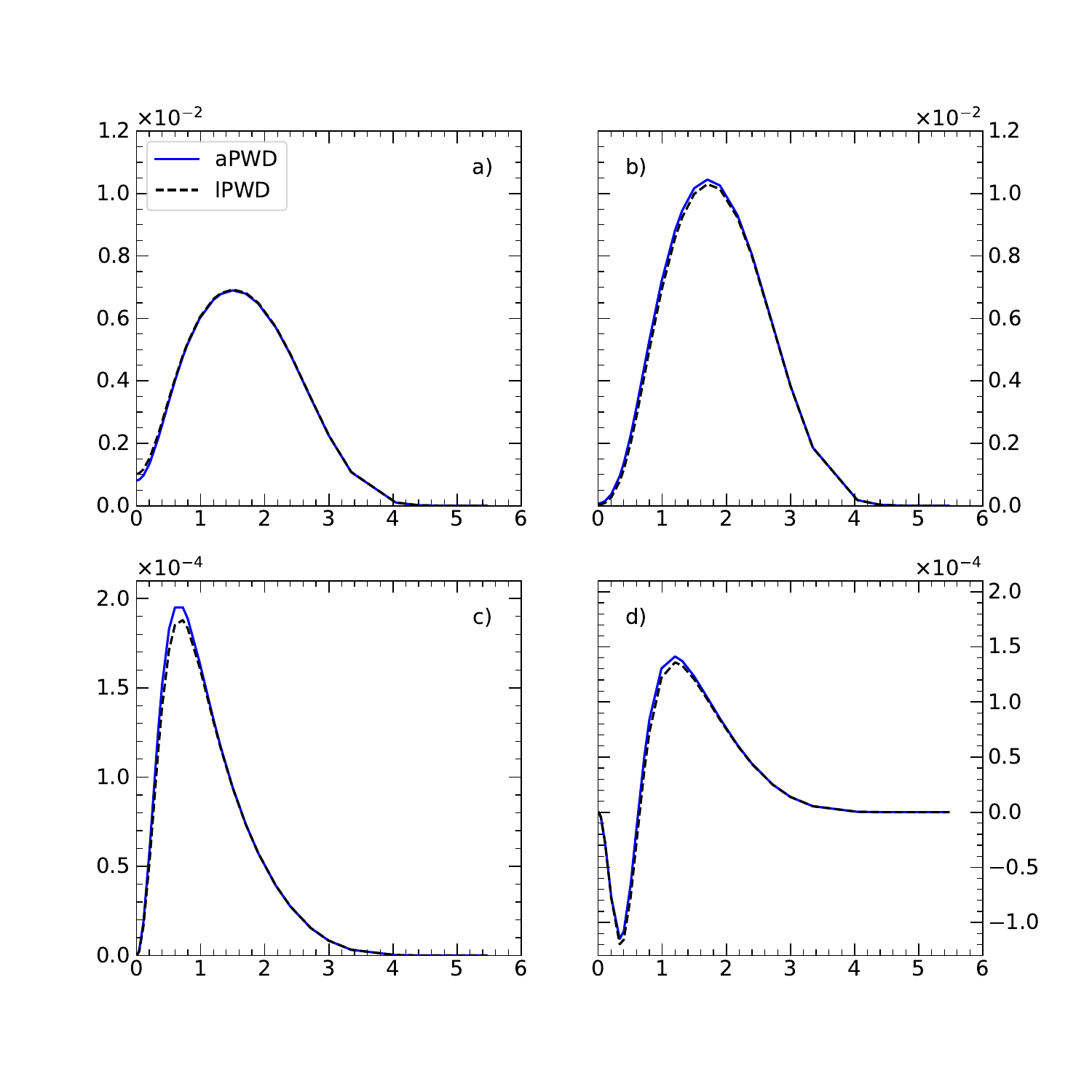}}
       \vspace{-1cm}
      \end{center}
                 \caption{  $2\pi$ (right panels) and $1\pi$  (left panels) $\mathrm{\Lambda NN}$ matrix elements  $\langle p^{\prime}_{12}=0.20550664,  q^{\prime}_{3}=0.20550664,  \alpha^{\prime}_\mathrm{\Lambda \mathrm{NN}} | V^{\mathrm{\Lambda NN}} | p_{12}, q_3 =0.20550664, \alpha_{\mathrm{\Lambda NN}} \rangle$, computed using aPWD (solid lines) and lPWD (dashed line),  as a  function of  $p_{12}$  in the $(J^{\pi},T)=(1/2^+,0)$ partial-wave state and for $(\alpha^{\prime}_{\mathrm{\Lambda NN}},\alpha_{\mathrm{\Lambda NN}})$: a), b) (1,1), c), d) (2,2). All matrix elements are in fm\textsuperscript{5} and have been regularized with a cutoff of $\Lambda=550$ MeV }
    \label{fig:2Pe-1Pe_VLNN}
         \end{figure}  
\begin{figure}[tbp] 
      \begin{center}
        { \includegraphics[width=0.5\textwidth]{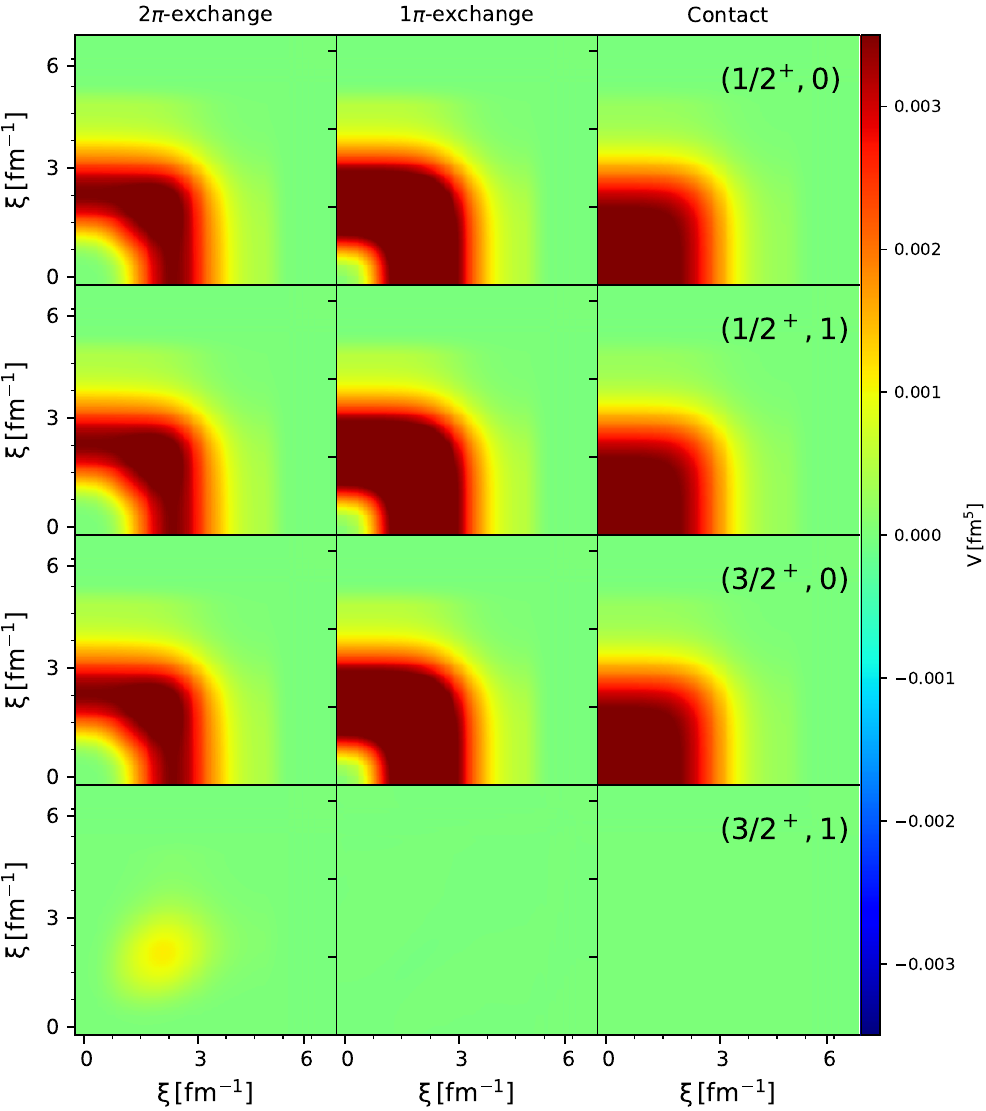}}
      \end{center}
                 \caption{ Matrix elements of  2$\pi$-,  1$\pi$- and contact-  $\mathrm{\Lambda NN}$ potentials  $\langle p^{\prime}_{12} q^{\prime}_{3} \alpha^{\prime}_\mathrm{\Lambda NN} | V_\mathrm{\Lambda NN} | p_{12} q_3 \alpha_\mathrm{\Lambda NN} \rangle$  as a function of the hypermomentum $\xi^2 = p^2_{12} + \frac{3}{4}q_3^2$ at a hyperangle $\tan \theta = p_{12}/(\sqrt{3}/2q_3) = \frac{\pi}{4}$ and in different partial-wave states  with $(J^{\pi}, T)= (1/2^+,0), (1/2^+,1), (3/2^+,0), (3/2^+,1), $ All matrix elements are in fm\textsuperscript{5} and have been regularized with a cutoff of $\Lambda=550$ MeV. The matrix elements in the partial wave state $(J^{\pi}, T)= (3/2^+,1)$ have been multiplied by a factor of 10 in order to make them more visible.  }
    \label{fig:2Pe-1Pe-CT_VLNN}
         \end{figure}  
         
 Table~\ref{tab:compare_VYNNmatrix1pect} lists the $1\pi$-exchange and contact $\mathrm{\Lambda NN}$ matrix elements in the partial-wave state with $(J^{\pi},T) = (1/2^+, 0)$. The LECs in Eqs.~(\ref{eq:LNNct}, \ref{eq:LNNope}) are set to 
 $D^{\prime}_1 =0$, $D^{\prime}_2 = \frac{2 C }{9 f_0^2 \Delta}  = 0.6268$ fm\textsuperscript{3} with $C=3/4\, g_A=0.9525$,  $f_0 = 93\,$MeV,   $ \Delta =300\,$MeV
 and 
 $C^{\prime}_2=0$,
 $C^{\prime}_1 = C^{\prime}_3 = \frac{1}{72 f_0^4 \Delta} = 0.1852\,$fm\textsuperscript{5} within the so-called decuplet saturation scheme. We do not show here the results in the $(J^{\pi},T) = (3/2^+, 0)$ state but stress that similar agreement of better than $0.5\%$ is also observed for the $1\pi$-exchange and contact potential  matrix elements in this partial-wave state. Fig.~\ref{fig:2Pe-1Pe_VLNN} shows the aPWD and lPWD $1\pi$- and $2\pi$-exchange  $\mathrm{\Lambda NN}$ 
 matrix elements,  \\
 $\langle p^{\prime}_{12},  q^{\prime}_{3},\alpha^{\prime}_\mathrm{\Lambda NN} | V^\mathrm{\Lambda NN} | p_{12}, q_3, \alpha_\mathrm{\Lambda NN} \rangle$ 
 in the partial-wave state $(1/2^+,0)$, as a function of the momentum $p_{12}$ for 
 $p^{\prime}_{12}=q^{\prime}_{3}=q_3 =0.20550664$~fm$^{-1}$. Note that
 the matrix elements in Fig.~\ref{fig:2Pe-1Pe_VLNN} have been regularized employing   a non-local regularization function of the form $f_{\Lambda}(p_{12},q_3) =\exp(-(p^2_{12} + \frac{3}{4} q_3^2)^2/\Lambda^4)$ with a cutoff of $\Lambda=550$~MeV. Such a non-local regularization function has the advantage that it does not depend on the angles and therefore can be applied to the potential independently of the partial-wave decomposition. 
 The so-called semi-local momentum-space (SMS) regularization developed by the Bochum group has however shown some advantages over the non-local regularization for the case of NN and 3NF forces \cite{Epelbaum:2022cyo}. The application of the SMS regularization to chiral YNN forces will be studied in \cite{HL:2024}. Finally, Fig.~\ref{fig:2Pe-1Pe-CT_VLNN} displays the $2\pi$-, $1\pi$-exchange and contact $\mathrm{\Lambda NN}$ matrix elements, \newline
  $\langle p^{\prime}_{12} q^{\prime}_{3} \, \alpha^{\prime}_{\mathrm{\Lambda NN}}=1 | V^{\mathrm{\Lambda NN}}  | p_{12} q_{3} \,\alpha_\mathrm{\Lambda\mathrm{NN}}=1\rangle$, in several par\-tial-wave states $(J^{\pi}, T) = (1/2^+,0), (1/2^+,1), (3/2^+,0)$ and $(3/2^+,1)$ as a function
  of the hyperspherical coordinate $\xi^2 = p^2_{12} + \frac{3}{4}q_3^2$  and  at a hyperangle  $\tan \theta = p_{12}/(\sqrt{3}/2q_3) = \frac{\pi}{4}$. Also here the non-local regularization function with a cutoff of $\Lambda=550$ MeV has been applied to all potentials.    The $V^{\mathrm{\Lambda NN}}$ matrix elements in the $(3/2^+,1)$ state have been scaled by a factor of 10 in order to make them visible on the plot. In general, the matrix elements in the higher partial-wave states that are not shown in Fig.~\ref{fig:2Pe-1Pe-CT_VLNN} are of at least two order of magnitude smaller than the ones in the $(1/2^+,0)$ state.
 
\section{\texorpdfstring{$A=3-5$}{A=3-5} hypernuclei with chiral \texorpdfstring{$\mathrm{\Lambda}$NN}{LambdaNN} }\label{sec:Eseparation_3_5}
In this section, we will investigate the possible contributions
of the chiral $\mathrm{\Lambda}$NN potentials to the separation energies of $A=3-5$ hypernuclei. As one can see from
Eqs.~(\ref{eq:LNNct}-\ref{eq:LNNtpe}), 
the $\mathrm{\Lambda}$NN potential is characterized by 
five LECs ($C'_1-C'_3$, $D'_1$, $D'_2$) which are difficult to determine due to the scarcity of the experimental data. However, 
using the decuplet saturation approximation the LECs can be qualitatively estimated. Specifically, they can be expressed
in terms of contact interactions for $BB\to BB^*$,
with pertinent LECs denoted by $H_i$ in 
Ref.~\cite{Petschauer:2016pbn}.
Then we are left with only one unknown LEC ($H'=H_1+3H_2$) for the 
case of $V_\mathrm{\Lambda NN}$  (and 
two LECs when both $\mathrm{\Lambda}$NN and $\mathrm{\Sigma}$NN are considered)
\cite{Petschauer:2016pbn},
\begin{eqnarray}
\label{eq:decuplet}
&  C_{1}^{\prime} = C_{3}^{\prime} = \frac{H^{\prime 2}}{72 \Delta},  \qquad   &  C^{\prime}_2 =0, \cr
 &D^{\prime}_{1} =0,  \qquad &  D_2^{\prime} = \frac{2C H^{\prime}}{9 \Delta},\cr
 & 3b_0 + b_D =0, \qquad  & 2b_2 + 3b_4 =-\frac{C^2}{\Delta}.
\end{eqnarray}
Here $\Delta$ is the decuplet-octet baryon mass difference and
$C = 3/4\, g_A \approx 1$ is the $B^*B\phi$ coupling constant
\cite{Petschauer:2016pbn}. 
As evidenced by Eq.~(\ref{eq:decuplet}),
decuplet saturation fixes also the sub-leading 
meson-baryon LECs, i.e. the $b_i$. Note, however, that within decuplet saturation some LECs are zero and thus the most general
structure of the YNN forces is not explored.

In general, the remaining LEC is expected to be  determined via a fit to the binding energies of the s-shell hypernuclei, which is beyond the scope of this study and will be thoroughly studied in a separate application \cite{HL:2024}. For our purpose of exploring the chiral $\mathrm{\Lambda}$NN interactions here, it is sufficient to assume a realistic scale for $H^{\prime}$. 
Therefore, we will adopt 
$H^{\prime} = 1/f^{2}_{0}$, as suggested in \cite{Petschauer:2016pbn} 
based on dimensional scaling arguments,
for all the calculations presented in this section. The separation energies for $A=3-5$ hypernuclei, computed using the two-body YN 
potential NLO19 with a cutoff of 
$\Lambda=550$ MeV
in combination with the \mbox{$2\pi$-,} $1\pi$-exchange and contact $\mathrm{\Lambda}$NN 
potentials, are listed in Table~\ref{tab:Energy_3_5}. The semi-local momentum-space (SMS) NN and 3N forces at 
$\mathrm{N^4LO^+}$ and  $\mathrm{N^2LO}$, likewise regularized with a cutoff 
of $\Lambda=550$~MeV, 
have been employed for describing the nuclear interaction. For $A=4$\,,$5$ hypernuclei also 3NFs contribute, for which we take the leading SMS regularized chiral 3NFs as specified for example in Table~1 of Ref.~\cite{Le:2023bfj}. 
For the calculations with the NCSM, all interactions have been evolved with the similarity renormalization group (SRG) at a flow parameter of 
$\lambda=1.88$~fm\textsuperscript{-1} up to 
induced 3BFs (in 3N, $\mathrm{\Lambda}$NN and 
$\mathrm{\Sigma}$NN). We have carefully checked 
that, for this flow parameter and using interactions up to the three-body level, 
the uncertainty due to omitted induced many-body forces is negligible (see also  \cite{Le:2023bfj}).   At the same time, NCSM calculations converge in reasonably sized model spaces \cite{LENPIC:2022cyu,Le:2022ikc,Le:2023bfj}. As discussed in the previous section, the chiral $\mathrm{\Lambda}$NN potential matrix elements at partial-wave states with the total angular momentum 
$J \ge 5/2$ are very small, their contributions to the binding energies are therefore  expected to be insignificant. Indeed, we  have observed that the $\mathrm{\Lambda}$NN 3BFs with $J =5/2$ contribute only a few keV to the binding energies in the $A=4,5$ systems. Therefore, for the calculations for $A \ge 4$ systems, the  $\mathrm{\Lambda}$NN matrix elements $V^{\mathrm{\Lambda NN}}$ with  $J \ge 7/2$ will be omitted,  whereas all the possible isospin  states $T =0,1,2$  and parities   are taken into account.

As already mentioned, for the $^3_{\mathrm{\Lambda}}\mathrm{H}$ system, we provide results from both the NCSM \cite{Le:2022ikc,Le:2023bfj} and the Faddeev approach \cite{Kamada:2023txx}. The 
 energies for the $A=4,5$ systems have been computed only
within the NCSM. Clear\-ly, the difference between the two $A=3$
results are smaller than the estimated uncertainty for the NCSM approach. 
The contribution of the contact potential \\ $V^{\mathrm{\Lambda NN}}_{ct}$ to $B_{\mathrm{\Lambda}}(^3_{\mathrm{\Lambda}}\mathrm{H})$ is negligibly small and repulsive, \\ whereas the $V^{\mathrm{\Lambda NN}}_{2\pi}$  and $V^{\mathrm{\Lambda NN}}_{1\pi}$ contributions are sizable and attractive,
amounting to about 70 and 40 keV, respectively.  Similarly, the effect of $V^{\mathrm{\Lambda NN}}_{ct}$ to the binding energy
$B_{\mathrm{\Lambda}}(^4_{\mathrm{\Lambda}}\mathrm{He}, 0^+)$
is  repulsive but with 30~keV rather insignificant. It becomes, however, moderately
repulsive in the $^4_{\mathrm{\Lambda}}\mathrm{He}(1^+)$ and
$^5_{\mathrm{\Lambda}}\mathrm{He}$ states, contributing about 50 and 200~keV, respectively. Interestingly, both the $1^+$ state in the $A=4$ system and $^5_{\mathrm{\Lambda}}\mathrm{He}$ are largely overbound with the $2\pi$- and $1\pi$-exchange $\mathrm{\Lambda}$NN potentials,
with respect to the present experimental 
information \cite{HypernuclearDataBase}, 
while the ground state in $A=4$ remains underbound. 
Since the sign of the LECs parameterizing the contact interaction is largely determined within the decuplet approximation, it could  be expected that the contribution from the contact terms  remains repulsive for any combinations of the 
$H_1$ and $H_2$ LECs. Note however that different choices for $H'$ allow for a partial cancelation of   $V^{\mathrm{\Lambda NN}}_{2\pi}$ and $V^{\mathrm{\Lambda NN}}_{1\pi}$.  In addition, let us also mention  that the inclusion of the chiral $\mathrm{\Sigma}$NN potentials does not change the behaviour  observed in these light hypernuclei. 
A careful analysis of the $H'$ or $H_1$ and $H_2$
dependence of the separation energies of the $s$-shell hypernuclei is beyond the scope of this 
work. 
 \begin{table*}[tbp]
\renewcommand{\arraystretch}{1.6}
\begin{center}
  \setlength{\tabcolsep}{0.2cm}
\begin{tabular}{l l r r r r r}
\hline
& &   w/o $\mathrm{\Lambda}$NN & w. $2\pi$-ex $\mathrm{\Lambda}$NN  &  w. $1\pi$-ex $\mathrm{\Lambda}$NN  & w. ct $\mathrm{\Lambda}$NN & Exp.~\cite{HypernuclearDataBase}\\
 \hline \hline
NCSM & {$^3_{\mathrm{\Lambda}}\mathrm{H}$} &  $0.080 \pm 0.006$   & $0.153 \pm 0.004$  & $0.121 \pm 0.005$ & $0.076  \pm 0.007  $  & $0.164 \pm 0.043$ \\

FY &  &   $0.087 $      & $0.152$  & $0.129$ & 
$0.080$ &\\ 
\hline
\hline
NCSM &
$^4_{\mathrm{\Lambda}}\mathrm{He}(0^+)$  &  $1.432 \pm 0.010$   &  $1.810 \pm 0.006$  & $ 1.619 \pm 0.007$ & $1.400 \pm 0.010$  & $2.347 \pm 0.036 $\\
& $^4_{\mathrm{\Lambda}}\mathrm{He}(1^+)$  &  $1.164 \pm 0.014$   &  $1.744 \pm 0.007$   & $1.427 \pm 0.009$   &  $1.117 \pm 0.016$  & $0.942 \pm 0.036$ \\
& $^5_{\mathrm{\Lambda}}\mathrm{He}$  &  $ 3.174 \pm 0.020$   &   $ 4.618 \pm 0.011$  &  $3.757 \pm 0.034 $  & $ 2.961 \pm 0.031$  & $3.102 \pm 0.030 $ \\
 \hline
  \end{tabular}
\end{center}
\caption{Separation energies for s-shell $\mathrm{\Lambda}$ hypernuclei without $\mathrm{\Lambda}$NN 3BF and with
$2\pi$-exchange, $1\pi$-exchange, or contact
3BF. 
All calculations are based on the SMS N\textsuperscript{4}LO\textsuperscript{+}(550) and NLO19(550) potentials for NN and YN, respectively, and on chiral $\mathrm{\Lambda}$NN 3BFs with non-local 
regulator of $\Lambda =550$~MeV. For the NCSM calculations all potentials have been SRG-evolved at a flow parameter of $\lambda=1.88$~fm\textsuperscript{-1}.
Also, an uncertainty estimate for the results is provided.} 
\label{tab:Energy_3_5}
\renewcommand{\arraystretch}{1.4}
\end{table*}
\section{Conclusions}
In this work, we examined two different approaches, lPWD and aPWD, to efficiently perform the partial-wave decomposition
of three-body forces, for the chiral  
$\mathrm{\Lambda}$NN (YNN) interactions. 
The $\mathrm{\Lambda}$NN matrix elements of the two methods were compared with each other in detail. In general, an agreement of better than $0.1\%$ is observed for the $2\pi$-exchange potential, whereas the difference in all the $1\pi$-exchange and contact $\mathrm{\Lambda}$NN potentials  matrix elements is smaller than $0.5\%$. Such a benchmark provides a solid confirmation of the correctness of our implementations and is of importance for any future calculations 
that include the chiral YNN 3BFs. 

As first application, we explored 
the possible impact of the leading chiral $\mathrm{\Lambda}$NN
potential on the separation energies of light hypernuclei. The 
 sub-leading meson-baryon LECs appearing in the 2$\pi$-exchange 
 3BF and the LECs in the 1$\pi$-exchange contribution and the
 six-baryon contact term were estimated via decuplet saturation 
 and assuming values for the LECs based on 
 dimensional scaling arguments. 
 It turned out that the weakly repulsive $\mathrm{\Lambda}$NN contact interaction leads to a small contribution to the binding energies in all $A=3-5$
hypernuclei, whereas the two other contributions, $V^{\mathrm{\Lambda NN}}_{2\pi}$ 
 and $V^{\Lambda\mathrm{NN}}_{1\pi}$, are
 moderately attractive for our choice of the
 only remaining, LEC $H'$. The size of the individual contributions are significant even 
 for $^3_\Lambda$H. This is somewhat surprising 
 since estimates for chiral N$^2$LO contributions so far indicated negligible 
 $\Lambda$NN force contributions \cite{Haidenbauer:2019boi, Le:2023bfj}. 
But the case studied here also leads to overbinding for  the $J^\pi=1^+$ state of $^4_{\mathrm{\Lambda}}\mathrm{He}$ and $^5_{\Lambda}\mathrm{He}$ while $^4_{\Lambda}\mathrm{He}(0^+)$ is still clearly underbound. The interesting question whether one can determine an optimal combination of the LECs within the decuplet approximation so that all light hypernuclei are well described, should be and will be addressed in a future study. 
In such a study, it should also be addressed whether the $\Lambda$NN force contribution to 
$^3_\Lambda$H remains sizable. 

\begin{acknowledgements}
HL, JH, UGM and AN  thank Stefan Petschauer for collaboration at the early stage of this work. This project is part of the ERC Advanced Grant ``EXOTIC'' supported the European Research Council (ERC) under the European Union’s Horizon 2020 research and innovation programme (grant agreement No. 101018170). This work is further supported in part by the Deutsche Forschungsgemeinschaft (DFG, German Research Foundation) and the NSFC through the funds provided to the Sino-German Collaborative Research Center TRR110 ``Symmetries and the Emergence of Structure in QCD'' (DFG Project ID 196253076 - TRR 110, NSFC Grant No. 12070131001),
and by the MKW~NRW under the funding code NW21-024-A.
The work of HK, MK and  KM is supported by Japan Society for the Promotion of Science (JSPS) KAKENHI Grants No. JP19K03849 and No. JP22K03597.
The work of UGM was supported in part by The Chinese Academy
of Sciences (CAS) President's International Fellowship Initiative (PIFI)
(grant no.~2025PD0022). We also acknowledge support of the THEIA net-working activity 
of the Strong 2020 Project. The numerical calculations were performed on JURECA
of the J\"ulich Supercomputing Centre, J\"ulich, Germany.
\end{acknowledgements}

\bibliographystyle{unsrturl}

\bibliography{bibliography.bib}

\end{document}